\documentclass[a4paper, conference]{IEEEtran}
\IEEEoverridecommandlockouts
\usepackage{graphicx} 
\usepackage{amsmath,amssymb,latexsym}
\usepackage{lmodern}                   
\usepackage[T1]{fontenc}               
\usepackage{textcomp}                  
\usepackage{booktabs}
\usepackage{balance}
\usepackage{algorithmic}
\usepackage{algorithm}
\usepackage{txfonts}
\usepackage{subfigure}
\usepackage{cite}
\usepackage[section]{placeins}
\usepackage{float}
\usepackage{color}

\title{Nonparametric Variational Bayesian Learning for Channel Estimation with OTFS Modulation}
\author{\IEEEauthorblockN{Chong Cao\textsuperscript{*}, Zhuyu Liu\textsuperscript{\S}, Zheng Dong\textsuperscript{*}, Yong Zhou\textsuperscript{\dag},  and He (Henry) Chen\textsuperscript{\#}}
	\IEEEauthorblockA{\textsuperscript{*}School of Information Science and Engineering, Shandong University, China. Email:zhengdong@sdu.edu.cn.}
      \IEEEauthorblockA{\textsuperscript{\S}Aerospace System Engineering Shanghai, Shanghai, China. Email: liuzhuyu@163.com}
	\IEEEauthorblockA{\textsuperscript{\dag}School of Information Science and Technology, ShanghaiTech University, China. Email: zhouyong@shanghaitech.edu.cn}

	\IEEEauthorblockA{\textsuperscript{\#}Department of Information Engineering, the Chinese University of Hong Kong, Hong Kong SAR, China. \\Email:he.chen@ie.cuhk.edu.hk}  
	\thanks{{This research was supported in part by the National Key R\&D Program of China under Grant 2024YFF0727101 and the Shandong Provincial Natural Science Foundation under Grant ZR2023LZH003.}    
}
}
\begin{document}

\maketitle

\begin{abstract}
Orthogonal time frequency space (OTFS) modulation has demonstrated significant advantages in high-mobility scenarios in future 6G networks. However, existing channel estimation methods often overlook the structured sparsity and clustering characteristics inherent in realistic clustered delay line (CDL) channels, leading to degraded performance in practical systems. To address this issue, we propose a novel nonparametric Bayesian learning (NPBL) framework for OTFS channel estimation. Specifically, a stick-breaking process is introduced to automatically infer the number of multipath components and assign each path to its corresponding cluster. The channel coefficients within each cluster are modeled by a Gaussian mixture distribution to capture complex fading statistics. Furthermore, an effective pruning criterion is designed to eliminate spurious multipath components, thereby enhancing estimation accuracy and reducing computational complexity. Simulation results demonstrate that the proposed method achieves superior performance in terms of normalized mean squared error compared to existing methods.
\end{abstract}

\section{Introduction}
Next-generation wireless communication systems support reliable information exchange in highly dynamic environments, such as vehicle-to-everything (V2X) and space-air-ground integrated network (SAGIN)~\cite{imt2023}. In such environments, scatterers in the channel can move at a high speed, leading to dual-selectivity fading. Traditional orthogonal frequency division multiplexing (OFDM) techniques may be difficult to apply due to the destruction of orthogonality caused by the Doppler effect. To address this issue, a new modulation scheme called orthogonal time frequency space (OTFS) was proposed by the authors in~\cite{hadani2017orthogonal} for such doubly selective channels.

Accurate channel estimation is essential for realizing the full potential of OTFS systems. One of the key challenges in OTFS channel estimation is off-grid delay and Doppler where the resolution of delay and Doppler is limited by the OTFS frame and duration.
In~\cite{raviteja2019embedded}, a threshold method and embedded pilot mode are proposed for the estimation of channel coefficients with OTFS modulation, which has a large spectral overhead. In order to improve spectral efficiency, some compressed sensing methods have been proposed. More specifically, the authors in~\cite{zhao2020sparse} proposed a novel pilot mode that does not need protective pilots, where the channel estimation problem is transformed into a sparse signal recovery problem. This method reduces the pilot overhead, but it is limited to estimating only integer delay and Doppler shift indices. 
Later, a low-complexity fast greedy sparse recovery algorithm termed two choice hard thresholding pursuit (TCHTP) for delay-Doppler (DD) channel estimation in OTFS systems was proposed in~\cite{kumari2023two}. This method efficiently estimates sparse channels with on-grid delays and off-grid Dopplers, addressing the challenge of unknown sparsity levels by simultaneously estimating the channel and its sparsity. 
To further enhance the channel estimation accuracy in OTFS,~\cite{wei2022off} proposed a novel two-dimensional off-grid scheme that balances complexity and performance by decoupling delay and Doppler shift estimation within a compressed sensing-based model.
On this basis, the authors in~\cite{WangQianli2021OTFS} proposed an OTFS symbol decomposition design across different dimensions, processing received symbols in two parallel streams. These two branches respectively estimate offline parameters and channel matrices, thereby forming complementarity in representing received symbols. 
Following these researches,~\cite{qiu2024turbo} proposed a Turbo-IFSLA-VBI algorithm that offers a novel joint design of grid parameters and channel estimation for OTFS systems. Its key innovation lies in integrating a three-layer hierarchical structured sparsity model with a turbo framework that combines variational Bayesian inference and message passing. The method achieves superior performance, reduced computational complexity, and faster convergence.
Moreover, with the objective of diminishing computational complexity, the authors in~\cite{ZhouYanxi20240ffgrid, ShanYaru2024offgrid, LiXiangjun2025grid} implemented a grid evolution scheme, whereby the virtual DD domain grid is updated as a non-uniform grid rather than the fixed uniform virtual grid. This approach led to a substantial reduction in complexity whilst maintaining satisfactory estimation performance. 

Most existing channel estimation methods for OTFS systems rely on simplified channel models, which leads to performance degradation in practical deployments~\cite{gunturu2021performance}.
The CDL model provides a more accurate representation of real-world propagation environments with multi-cluster and multi-path characteristics, yet its high parameter estimation complexity has not been thoroughly investigated in OTFS systems. To address this gap, we introduce a nonparametric Bayesian approach integrated with variational sparse Bayesian learning, proposing a novel channel estimation framework tailored for CDL-type channels in OTFS. This framework aims to capture the inherent sparsity and statistical properties of practical channels, thereby improving estimation accuracy and overall system robustness.
The main contributions of this paper are summarized as follows:
\begin{itemize}
\item We introduce a stick-breaking process for automated estimation of paths and clusters. Specifically, a nonparametric stick-breaking prior is incorporated into the Bayesian framework to automatically infer the number of multipath components and assign each path to its corresponding cluster. 
\item An enhanced channel estimation accuracy is achieved by employing statistical characterization of clusters. The channel parameters within each cluster are modeled with Gaussian mixtures, allowing the algorithm to capture and exploit cluster-specific statistical properties. By leveraging these statistics, the proposed method significantly improves the estimation accuracy of channel parameters under various non-Gaussian fading conditions.
\item In our design, a novel pruning criterion based on the sparsity-promoting property of variational posteriors is developed to eliminate insignificant or virtual multipath components, which considerably reduces the computational complexity of the algorithm.
\end{itemize}
\section{System Model}
\begin{figure}
    \centering
\includegraphics[width=0.8\linewidth]{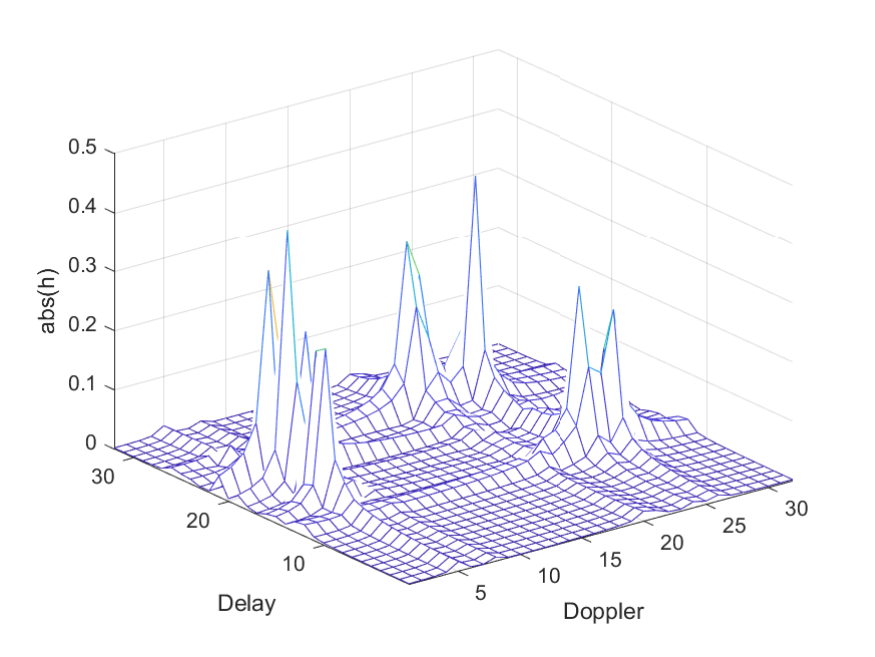}
    \caption{Channel in the DD domain.}
    \label{DDchanel}
\end{figure}

At the transmitter, $NM$ information symbols are arranged in the DD domain matrix $X_{DD} \in \mathbb{C}^{N \times M}$. The transmission signal has a time slot duration $T$, and the subcarrier spacing $ \Delta f $ satisfies $ \Delta f = \frac{1}{T} $.  Consider a DD domain channel with $ P $ propagation paths, each characterized by delay $ \tau_i $, Doppler shift $ \nu_i $, and channel gain $ h_i $, remaining quasi-static over a frame. For illustration, we depict the CDL channel in the DD domain in Fig.~\ref{DDchanel}.

We employ a clustered DD channel model which can better capture the key propagation characteristics of realistic wireless environments than simplified independent multipath models.
The central premise is that multipath components do not occur randomly but rather form distinct clusters, each corresponding to a dominant scatterer or a group of correlated scatterers in the environment. To closely emulate this physical reality, the channel is constructed using a Gaussian mixture model (GMM) framework for generating all cluster parameters. The center of each cluster, defined by its mean delay and mean Doppler value, randomly appears across the DD grid, reflecting the spatial distribution of scatterers. Within each cluster, individual paths exhibit delays and Doppler shifts that are distributed according to a Gaussian distribution around the cluster centroid, accurately modeling the diffusion effects caused by finite scatterer size and mobility.

By employing a Gaussian mixture based process for the channel generation, our model delivers high fidelity in mimicking the real-world channel behavior, including sparsity, clustering, and parameter dependency, thereby providing a realistic and robust foundation for evaluating OTFS. More specifically, we consider the following channel model:

\begin{equation}  
h(\nu, \tau) = \sum_{i=1}^{P} h_i \delta(\nu - \nu_i) \delta(\tau - \tau_i),  
\end{equation}  
where $\delta(\cdot)$ denotes the Dirac delta function.  
The input-output relationship in the DD domain:  
\begin{equation}  
y(k,l) = \sum_{k^\prime} \sum_{l^\prime}  h_w(l - l^\prime, k - k^\prime)x(l^\prime, k^\prime) + w(k,l),  
\label{DD_relation}  
\end{equation}  
where $h_w(k,l)$ denotes the $(k,l)$-th element in the effective impulse response matrix, and $x(k,l)$, $y(k,l)$, and $w(k,l)$ represent the transmitted signal, received signal, and noise terms at the $(k,l)$-th grid point, respectively. The term $h_w(k,l)$ is the effective impulse response, given by:  
\begin{equation}  
h_w(k - k^\prime, l - l^\prime) = \sum_{i=1}^{P} h_i w(\tilde{k}_i, \tilde{l}_i),  
\label{effective_response}  
\end{equation}  
where $\tilde{k}_i = k -  k^\prime - k_{\nu_i} $, $\tilde{l}_i = l - l^\prime - l_{\tau_i} $, and $k_{\nu_i} = \frac{\nu_i}{N \Delta f}$, $l_{\tau_i} = \frac{\tau_i}{M T} $ map the Doppler and delay shifts to the DD grid. The sampling function $w(\tilde{k}_i, \tilde{l}_i)$ decomposes into $w_\nu(\tilde{k}_i)$ and $w_\tau(\tilde{l}_i)$, such that  
\begin{equation}
\begin{aligned} 
w_\nu(\tilde{k}_i) &= \frac{1}{N} e^{-j \frac{(N-1)\pi \tilde{k}_i}{N}} \frac{\sin(\pi \tilde{k}_i)}{\sin(\pi \tilde{k}_i / N)},\\
w_\tau(\tilde{l}_i) &= \frac{1}{M} e^{j \frac{(M-1)\pi \tilde{l}_i}{M}} \frac{\sin(\pi \tilde{l}_i)}{\sin(\pi \tilde{l}_i / M)}.  
\end{aligned}
\end{equation}

\begin{figure}[ht]
        \centering        \includegraphics[width=1.\linewidth]{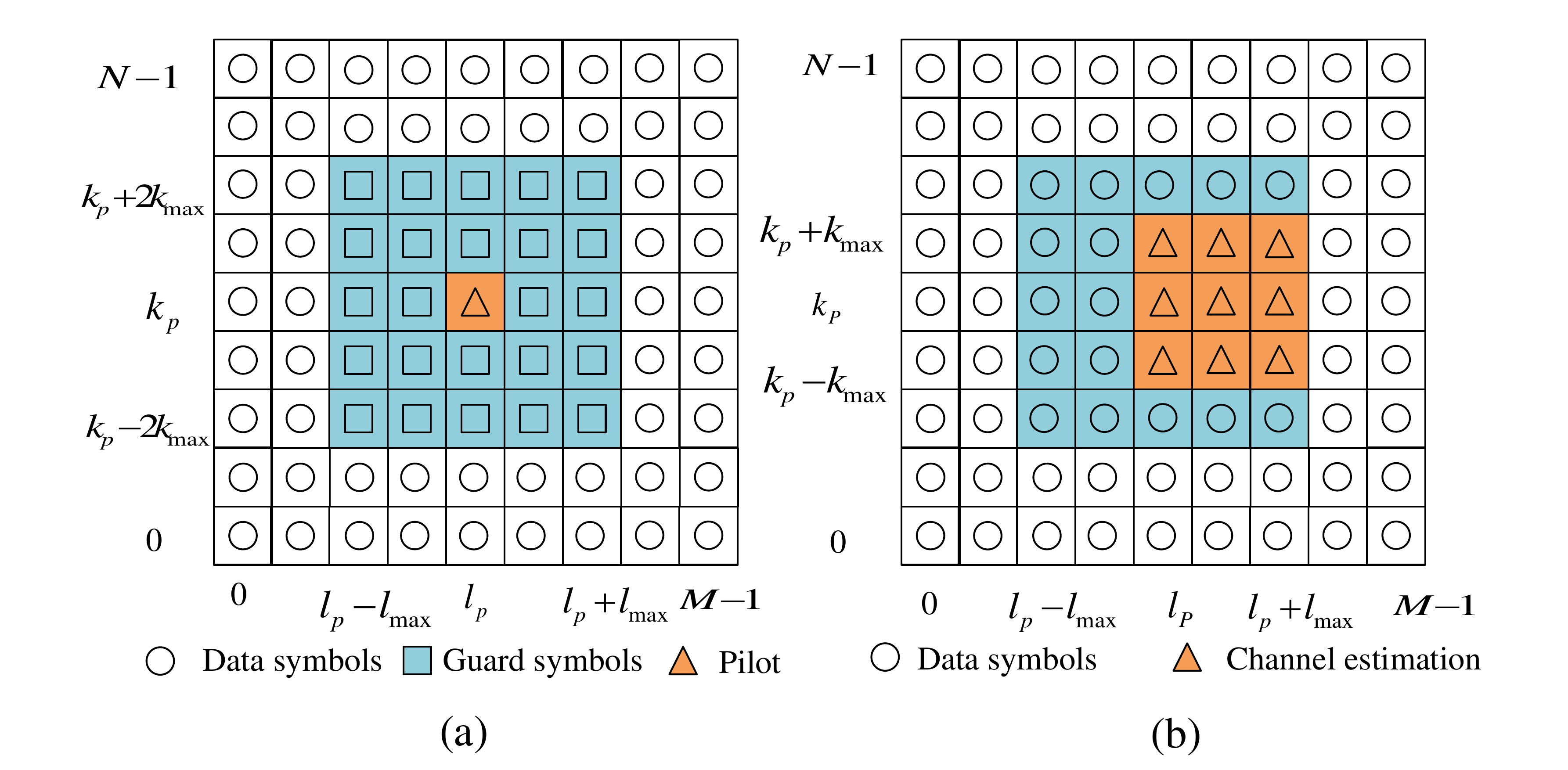}
        \caption{Tx and Rx symbols in the DD domain.}
        \label{frame}
    \end{figure}

The arrangement of the transmitted frame structure is illustrated in Fig.~\ref{frame}(a). 
The OTFS frame structure at the transmitter includes a pilot symbol $x_p$ at the $(k_p, l_p)$-th position surrounded by guard symbols. These guard symbols, which are set to zero, occupy the range  $y[k,l]$ within the range $l_p -l_{\max} \leq l \leq l_p + l_{\max}$ and $k_p - 2k_{\max} \leq k \leq k_p + 2k_{\max}$. 
At the receiver side, the corresponding frame structure is depicted in Fig.~\ref{frame}(b), where the received symbols $y[k, l]$ used for channel estimation are selected within the region defined by $l_p \leq l \leq l_p + l_{\max}$ and $k_p - k_{\max} \leq k \leq k_p + k_{\max}$.
    
It is worth noting that the subsequent sections omit the Data part of the frame and use the entire frame for channel estimation.

\section{Variational Nonparametric Bayesian Learning}
In this section, we first perform a successive linear approximation, then introduce a truncated stick-breaking process to model sparse channels, and finally outline the variational inference procedure for posterior approximation.
\subsection{Successive Linear Approximation}
The partial received signal in \eqref{DD_relation} for channel estimation~\cite{qiu2024turbo} is truncated as:  
\begin{equation}
\boldsymbol{y}_T = \boldsymbol{\Phi}(\boldsymbol{k}, \boldsymbol{l}) \boldsymbol{h} + \boldsymbol{w}_T,
\end{equation}
where $\boldsymbol{k}$ and $\boldsymbol{l}$ denote the vectors of Doppler and delay grid parameters, respectively. In practical systems, the limited resolution of the virtual Doppler and delay grids often leads to off-grid errors, where the true parameters $\boldsymbol{k}_{\nu}=[k_{\nu_1},k_{\nu_2},\cdots,k_{\nu_P}]^T$ and $\boldsymbol{l}_{\tau}=[l_{\tau_1},l_{\tau_2},\cdots,l_{\tau_P}]^T$ are non-integer and do not align perfectly with the predefined discrete grid points. To address this off-grid issue, a first-order linear approximation is applied to the sensing matrix $\boldsymbol{\Phi}(\boldsymbol{k}, \boldsymbol{l})$ around given points $(\boldsymbol{k}_{\nu}, \boldsymbol{l}_{\tau})$:
{
\begin{equation}
     \begin{aligned} 
    \boldsymbol{\Phi}(\boldsymbol{k}_{\nu}, \boldsymbol{l}_{\tau})& \approx \boldsymbol{\Phi}(\hat{\boldsymbol{k}_{\nu}}, \hat{\boldsymbol{l}_{\tau}}) + \boldsymbol{\Phi}_\nu(\hat{\boldsymbol{k}_{\nu}}, \hat{\boldsymbol{l}_{\tau}}) {\rm diag}(  \boldsymbol{k}_{\nu}-\hat{\boldsymbol{k}_{\nu}})\\   
    &+ \boldsymbol{\Phi}_\tau(\hat{\boldsymbol{k}_{\nu}}, \hat{\boldsymbol{l}_{\tau}}){\rm diag}(\boldsymbol{l}_{\tau} - \hat{\boldsymbol{l}_{\tau}}) \triangleq \bar{\boldsymbol{\Phi}}(\boldsymbol{k}_{\nu}, \boldsymbol{l}_{\tau}),
    \end{aligned}
\end{equation}
where the $\rm{diag(\cdot)}$ operator returns a diagonal matrix given a vector, or extracts the diagonal elements as a vector given a matrix. The $\boldsymbol{\Phi}_{\nu}(\hat{\boldsymbol{k}_{\nu}}, \hat{\boldsymbol{l}_{\tau}})\in \mathbb{C}^{(l_{\max}+1)(2k_{\max}+1) \times P}$ and $\boldsymbol{\Phi}_\tau(\hat{\boldsymbol{k}_{\nu}}, \hat{\boldsymbol{l}_{\tau}})\in \mathbb{C}^{(l_{\max}+1)(2k_{\max}+1) \times P}$ denote the first-order  
 gradients of $\boldsymbol{\Phi}(\hat{\boldsymbol{k}_{\nu}}, \hat{\boldsymbol{l}_{\tau}})\in \mathbb{C}^{(l_{\max}+1)(2k_{\max}+1) \times P}$, respectively, such that
 \begin{equation}
 \begin{aligned}
     &\boldsymbol{\Phi}_{\nu}(\hat{\boldsymbol{k}_{\nu}}, \hat{\boldsymbol{l}_{\tau}})=\frac{\partial \boldsymbol{\Phi}(\hat{\boldsymbol{k}_{\nu}}, \hat{\boldsymbol{l}_{\tau}})}{\partial \hat{\boldsymbol{k}_{\nu}}^T},\\
     &\boldsymbol{\Phi}_\tau(\hat{\boldsymbol{k}_{\nu}}, \hat{\boldsymbol{l}_{\tau}})=\frac{\partial \boldsymbol{\Phi}(\hat{\boldsymbol{k}_{\nu}}, \hat{\boldsymbol{l}_{\tau}})}{\partial \hat{\boldsymbol{l}_{\tau}}^T}.
  \end{aligned}
  \end{equation}
We update $\boldsymbol{\Phi}(\hat{\boldsymbol{k}_{\nu}}, \hat{\boldsymbol{l}_{\tau}})$ iteratively with the latest estimation results to ensure highly accurate estimates can be obtained.}

\subsection{Truncated Stick Breaking Process}
We assume that the maximum truncation level is $T$, and $\boldsymbol{C}\in \mathbb{Z}_{\{0,1\}}^{P\times T}$ is a $(0,1)$ matrix in which $c_{i,t}=1$ means that the $i$-th path corresponds to the $t$-th Gaussian component. $\theta_t$ denotes the probability that $c_{i,t}=1$, generated by the stick-breaking process 
\begin{equation}
\begin{aligned}
V_t&\sim Beta(\lambda_1,\lambda_2),\\
\theta_t&=V_t\prod^{t-1}_{\ell=1}(1-V_{\ell-1}),
\end{aligned} 
\end{equation}
where the stick-breaking weights ${\theta_t}$ satisfy $\sum_{t=1}^T \theta_t = 1$. 
This construction provides a finite-dimensional approximation to the Dirichlet process mixture model, facilitating tractable variational inference. The variational posterior distribution $q(\boldsymbol{V}, \boldsymbol{C})$ is factorized as:
\begin{equation}
    q(\boldsymbol{V}, \boldsymbol{C}) = \prod_{t=1}^{T-1} q(V_t; \lambda_{1,t},\lambda_{2,t}) \prod_{i=1}^P q(\boldsymbol{c}_i; \boldsymbol{r}_i),
\end{equation}
where $\boldsymbol{r}_i=[r_{i,1},r_{i,2},\cdots,r_{i,T}]^T$ are variational parameters for the multinomial assignment.

\subsection{GMM Model for Channel Coefficient Estimation}
\begin{figure}[ht]
    \centering
    \includegraphics[width=0.8\linewidth]{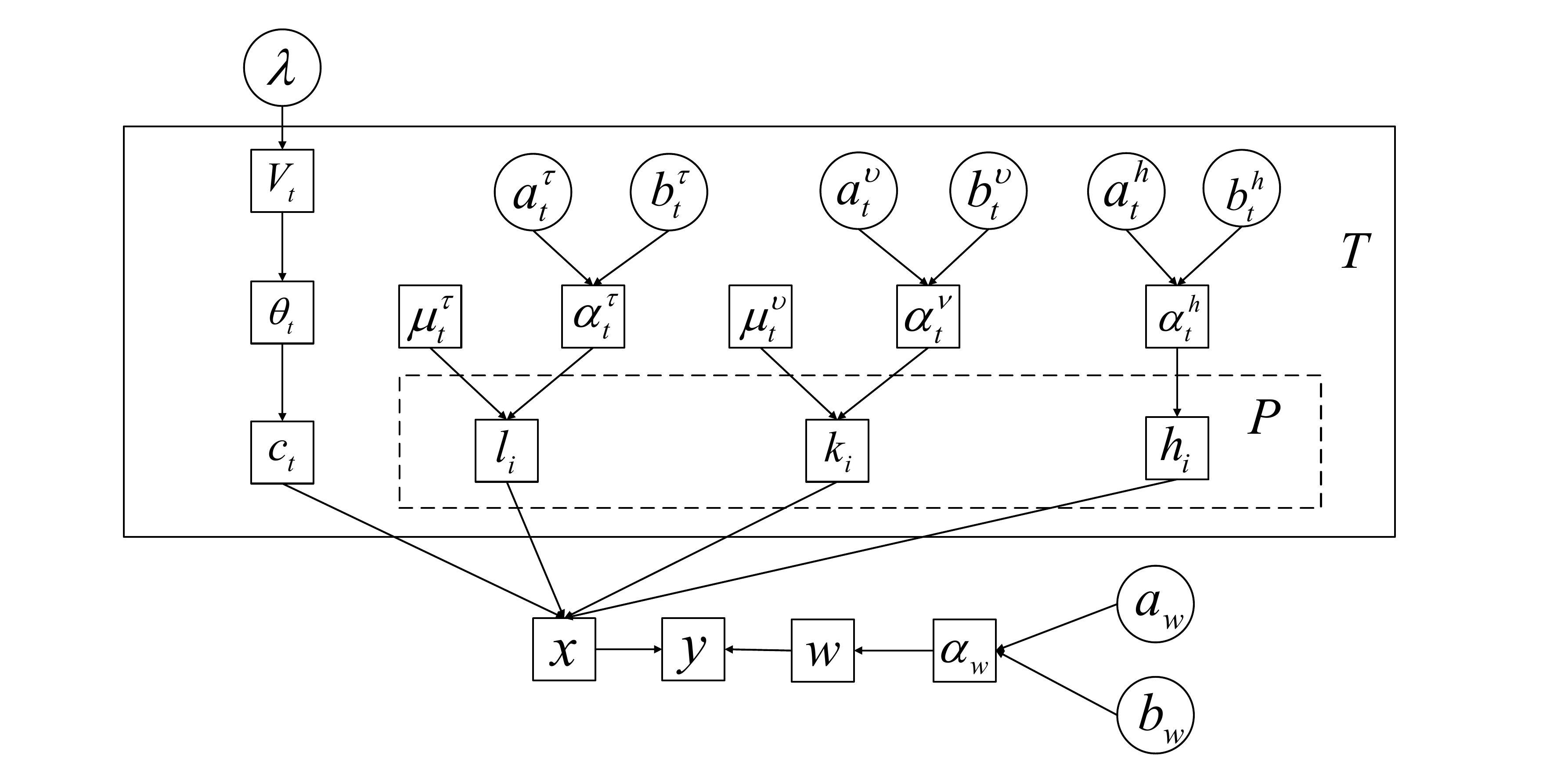}
    \caption{Graphical model for the proposed algorithm.}
    \label{factor}
\end{figure}

A complex Gaussian prior distribution is assigned to $\boldsymbol{h}$, and we model the sparse channel in the DD domain as follows:
	\begin{equation}
		p({\boldsymbol{h}}\mid \boldsymbol{C}, \boldsymbol {\alpha }^h)=\prod_{i=1}^{P}\prod_{t=1}^{T}{\mathcal{CN}}(h_i\mid 0, \alpha^h_{t})^{c_{i,t}},
	\end{equation}
where $\alpha^h_{t}$ is the precision parameter (i.e., reciprocal of the variance) of $h_i$.  We assign the Gamma distribution to $\alpha^h_{t}$ as the prior distribution
	\begin{equation}
		\begin{aligned}
			p(\alpha^h_{t};a^h_{t},b^h_{t})=&\Gamma \left ({{\alpha^h_{t}\mid a^h_{t},b^h_{t}} }\right).
		\end{aligned}
	\end{equation}
All $h_i$ that belong to the class $t$ share a common precision parameter, $\alpha^{h}_{t}$. 

Next, we assume the noise in~\eqref{DD_relation}  follows a white Gaussian distribution with mean $0$ and precision  $\alpha_w$, such that
	\begin{equation} 
		p\left(\boldsymbol{w} \mid \alpha _{w}\right) = \mathcal{ CN}\left (\boldsymbol{w}\mid 0,\alpha _{w}^{-1}\boldsymbol{I} \right).
	\end{equation}
    Then, we model  $\alpha_w$ as Gamma distribution with hyperparameters $a_w$ and $b_w$,
	\begin{equation}
		p({\alpha _{w}};a_w,b_w) = \Gamma ({\alpha _{w}}\mid a_w,b_w),
	\end{equation}
The prior distribution of $\boldsymbol k_{\nu}$ is given by:
        \begin{equation}
		\begin{aligned}
		  &p({\boldsymbol{k}_{\nu}}\mid \boldsymbol{C}, \boldsymbol{\mu}^{\nu}, \boldsymbol{\alpha_k 
            })=\prod_{i=1}^{P}\prod_{t=1}^{T}{\mathcal{N}}(k_{\nu_i} \mid \mu^{\nu}_t, \alpha^{\nu}_{t})^{c_{i,t}},
		\end{aligned}
        \end{equation}
    where $\boldsymbol{\mu}^{\nu}=[\mu^{\nu}_{1},\mu^{\nu}_{2},\cdots,\mu^{\nu}_{T}]^T$ is assigned a flat prior. All $k_{\nu_i}$ that belong to the class $t$ share a common precision parameter, $\alpha^{\nu}_{t}$. We model  $\alpha^{\nu}_{t}$ as Gamma distribution with hyperparameters $a^{\nu}_{t}$, $b^{\nu}_{t}$ 
	\begin{equation}
		p(\alpha^{\nu}_{t};a^{\nu}_{t},b^{\nu}_{t}) = \Gamma (\alpha^{\nu}_{t}\mid a^{\nu}_{t},b^{\nu}_{t}).
	\end{equation}
Now, the likelihood function can be given by
	\begin{equation} 
		p({\boldsymbol{y}_T}\mid{\boldsymbol{\Phi }},{\alpha_w}) = {\mathcal{ CN}}\left ({{{\boldsymbol{y}_T}\mid\boldsymbol{\Phi} \boldsymbol{h},{\alpha _{w}}^{ - 1}{\boldsymbol{I}}} }\right).
	\end{equation}
We use $\boldsymbol{\Omega} = \{ {\boldsymbol{h}},\,  \boldsymbol {k}_{\nu},\, \boldsymbol {l}_{\tau},\, \boldsymbol {\alpha }_h,\, \boldsymbol {\alpha }_k,\, \boldsymbol {\alpha }_l,\, \alpha _{w},\, \boldsymbol\mu_h,\, \boldsymbol\mu_l, \, \boldsymbol\mu_k,\, \boldsymbol\lambda_{1},\, \boldsymbol\lambda_{2}, \,\boldsymbol r, \, \boldsymbol{C}\}$ to denote the collection of all model parameters, and the joint probability density function (PDF) can be expressed as
    \begin{equation}
        \begin{aligned}
            p({\boldsymbol{y}_T},\boldsymbol{\Omega})& =
            p({\boldsymbol{y}_T}\mid \boldsymbol{\Phi }, \boldsymbol{h }, {\alpha _{w}})
            p(\alpha_w)\\  
            &\times \prod^P_{i=1}\prod^T_{t=1}\bigg[
            p(h_i\mid \alpha^h_{t})
            p(k_i\mid \mu^{\nu}_{t},\alpha^{\nu}_{t})
            p(l_i\mid \mu^{\tau}_{t},\alpha^{\tau}_{t})\bigg]^{c_{i,t}}\\  
            &\times \prod^T_{t=1}p(\alpha^h_{t};a^h_{t},b^h_{t})
            p(\alpha^{\nu}_{t};a^{\nu}_{t},b^{\nu}_{t})
            p(\alpha^{\tau}_{t};a^{\tau}_{t},b^{\tau}_{t})\\
            &\times \prod^{P}_{i=1}P(\boldsymbol{C}_i\mid V_t)
            \prod^{T}_{t=1}P(V_t\mid 1,\lambda).
            \end{aligned}
    \end{equation}
The graphical model of the factors is illustrated in Fig.~\ref{factor}.
    
 \subsection{Variational Bayesian Inference}

	We use the method of variational inference to approximate the posterior distribution of the variables, which assumes that the joint posterior distribution can be decomposed as
	\begin{equation}
    \begin{aligned}
        p(\boldsymbol{\Omega} \mid {\boldsymbol{y}_T}) \approx 
        \prod\limits_{i} q \left( {{\omega_i}} \right){\text{d}}{\omega_i}.
    \end{aligned}
	\end{equation}
	where $\omega_i$ represents one of the variables in $\boldsymbol{\Omega}$. By assuming the variables are independent of each other, we maximize the lower bound of the likelihood function, which allows us to express the logarithm of the approximate posterior distribution for each variable as 
        \begin{align*}
        	\ln q\left( {{\omega_i}} \right)& = \int {\ln }  p({\boldsymbol{y}_T}, \boldsymbol{\Omega})\prod\limits_{\ell \ne i} q \left( {{\omega_\ell}} \right){\text{d}}{\omega_\ell} \nonumber+ {\text{const}},
        \end{align*}
where “const" represents a constant term independent of $\omega_i$. The iterative update formula for each parameter is derived as follows.
 
The variational parameters $\tilde \lambda_1$ and $\tilde \lambda_2$ of the Beta distribution for the stick-breaking variable $V_t$ are formulated as
\begin{equation}
    \tilde \lambda_{1,t}=\lambda_1+\sum_{i=1}^P r_{i,t},\quad
    \tilde \lambda_{2,t}=\lambda_2+\sum_{i=1}^P\sum_{\ell=t}^T r_{i,\ell}.
    \label{ladmaitr}
\end{equation}
Then, $\boldsymbol{C}$ is determined by $c_{i,t}=r_{i,t}$, where $r_{i,t}$ denotes the variational expectation that the $i$-th path belongs to the $t$-th Gaussian component, and it is derived as
\begin{align}
        r_{i,t}&\propto \exp\bigg[ \psi(\tilde a^h_{i,t})-\ln{\tilde b^h_{i,t}}+\frac{1}{2}\bigg( \psi(\tilde a^{\nu}_{i,t})-\ln{\tilde b^{\nu}_{i,t}}+\psi(\tilde a^{\tau}_{i,t})-\ln{\tilde b^{\tau}_{i,t}} \bigg)\nonumber\\
        &-\frac{a^h_{i,t}}{b^h_{i,t}}\bigg( |\hat h_i|^2+{\sigma_{i,t}^h}^2 \bigg)-\frac{a^{\nu}_{i,t}}{2b^{\nu}_{i,t}}\bigg( \hat k_i^2+{\sigma_{i,t}^{\nu}}^2 +{\mu_{i,t}^{\tau}}^2-2\hat k_{\nu_i}\hat\mu_{i,t}^{\nu} \bigg)\nonumber\\
        &-\frac{a^{\tau}_{i,t}}{2b^{\tau}_{i,t}}\bigg( \hat l_i^2+{\sigma_{i,t}^{\tau}}^2 +{\mu_{i,t}^{\tau}}^2-2\hat l_{\nu_i}\hat\mu_{i,t}^{\tau} \bigg)+\psi(\tilde \lambda_{1,t})-\psi(\tilde \lambda_{1,t}+\tilde \lambda_{2,t})\nonumber\\
        &+\sum_{\ell=1}^{t-1}\bigg( \psi(\tilde \lambda_{1,\ell})-\psi(\tilde \lambda_{1,\ell}+\tilde \lambda_{2,\ell}) \bigg) 
        \bigg],
        \label{ritr}
\end{align}
in which $\psi(\cdot)$ is digamma function~\cite{blei2006variational}, $\hat{\cdot}$ denote the estimates from the previous iteration.
 
The covariance matrix $\boldsymbol{\Sigma}_h$ and mean vector $\boldsymbol {\mu}_h$ are formulated as
\begin{equation}
\begin{aligned}
\boldsymbol{\Sigma}_h &= \bigg[ \langle \alpha_w \rangle \langle\boldsymbol{\Phi}^H(\boldsymbol{k}_{\nu},\boldsymbol{l}_{\tau})\boldsymbol{\Phi}(\boldsymbol{k}_{\nu},\boldsymbol{l}_{\tau})\rangle_{\boldsymbol{k}_{\nu},\boldsymbol{l}_{\tau}} 
 + \operatorname{diag} \big\langle \boldsymbol{\gamma}^h \big\rangle \bigg]^{-1} \\
\boldsymbol{\mu}_h &= \langle \alpha_w \rangle \boldsymbol{\Sigma}_h \big\langle {\boldsymbol{\Phi}}^H(\boldsymbol{k}_{\nu}, \boldsymbol{l}_{\tau}) \big\rangle_{\boldsymbol{k}_{\nu}, \boldsymbol{l}_{\tau}} \boldsymbol{y}_T,
\end{aligned}
\label{hitr}
\end{equation}
  where $\langle \cdot  \rangle$ is the expectation operation, the posterior expectation of $\alpha _{w}$ is $\langle \alpha_{w} \rangle =\tilde{a}_w/\tilde{b}_w$.
  For simplicity of subsequent derivations, we introduce the following two definitions for terms: 
    \begin{equation}
    \begin{aligned}                \langle{\boldsymbol{\Phi}}^H(\boldsymbol{k}_{\nu},\boldsymbol{l}_{\tau})\rangle_{\boldsymbol{k}_{\nu},\boldsymbol{l}_{\tau}}
    &\approx \langle\bar{\boldsymbol{\Phi}}^H(\boldsymbol{k}_{\nu}, \boldsymbol{l}_{\tau})\rangle_{\boldsymbol{k}_{\nu},\boldsymbol{l}_{\tau}}\\
    &= \boldsymbol{\Phi}^H(\hat{\boldsymbol{k}}_{\nu}, \hat{\boldsymbol{l}}_{\tau}),\\
    \langle\boldsymbol{\Phi}^H(\boldsymbol{k}_{\nu},\boldsymbol{l}_{\tau})\boldsymbol{\Phi}(\boldsymbol{k}_{\nu},\boldsymbol{l}_{\tau})\rangle_{\boldsymbol{k}_{\nu},\boldsymbol{l}_{\tau}}
        &\approx \langle\bar{\boldsymbol{\Phi}}^H(\boldsymbol{k}_{\nu}, \boldsymbol{l}_{\tau})\bar{\boldsymbol{\Phi}}(\boldsymbol{k}_{\nu}, \boldsymbol{l}_{\tau})\rangle \\
        &=\bar{\boldsymbol{\Phi}}^H(\hat{\boldsymbol{k}}_{\nu}, \hat{\boldsymbol{l}}_{\tau})\bar{\boldsymbol{\Phi}}(\hat{\boldsymbol{k}}_{\nu}, \hat{\boldsymbol{l}}_{\tau})\\
        &+\boldsymbol{\Phi}_{\nu}^H(\hat{\boldsymbol{k}}_{\nu}, \hat{\boldsymbol{l}}_{\tau})\boldsymbol{\Phi}_{\nu}(\hat{\boldsymbol{k}}_{\nu}, \hat{\boldsymbol{l}}_{\tau}) \odot\boldsymbol{\Sigma}_{\nu}\\
        &+\boldsymbol{\Phi}_{\tau}^H(\hat{\boldsymbol{k}}_{\nu}, \hat{\boldsymbol{l}}_{\tau})\boldsymbol{\Phi}_{\tau}(\hat{\boldsymbol{k}}_{\nu}, \hat{\boldsymbol{l}}_{\tau})\odot\boldsymbol{\Sigma}_{\tau}\\
        &\triangleq\boldsymbol{H}_h,
    \end{aligned}
    \end{equation}
    where $\boldsymbol{\Sigma}_{\nu}$ and $\boldsymbol{\Sigma}_{\tau}$ are the covariance matrices of $\boldsymbol{k}_{\nu}$ and $\boldsymbol{l}_{\tau}$, respectively, and $\odot$ is the Hadamard product. Moreover, $\boldsymbol {\gamma}^h=[{\gamma}^h_1,{\gamma}^h_2,\cdots,{\gamma}^h_P]^T$,  in which ${\gamma}^h_i$ is the precision coefficient of $h_i$,
  \begin{equation}
      \langle{\gamma}^h_i\rangle=\sum_{t=1}^T r_{i,t}\langle\alpha^h_t\rangle,
      \label{alphahitr}
  \end{equation}
   and $\langle\alpha^h_{t} \rangle=\tilde{a}_{t}/\tilde{b}_{t}$, respectively. The shape parameter $\tilde{a}_{t}^h$ and rate parameter $\tilde{b}_{t}^h$ are given by 	
    \begin{equation}
		{\tilde a_t^h} = a_t^h + \sum_{i=1}^P r_{i,t},\quad {\tilde b_t^h} = b_t^h + \sum_{i=1}^P r_{i,t}\left\langle {{{\left| {{h _i}} \right| }^2}} \right\rangle ,
        \label{alphaitr}
    \end{equation}
    where $\langle\mid h_{i}\mid^2\rangle=\mid\mu_{i}^h\mid^2+\boldsymbol{\Sigma}_{i,i}^h$,  $\mu_{i}^h$ and $\boldsymbol{\Sigma}_{i,i}^h$ are the $i$-th entry of $\boldsymbol {\mu}^h$ and the $i$-th diagonal entry of $\boldsymbol {\Sigma}^h$, respectively.
The shape parameter $\tilde{a}_w$ and rate parameter $\tilde{b}_w$ are derived as
    \begin{equation}
    \begin{aligned}
        &\tilde a_w = a_w +(l_{\max}+1)(2k_{\max}+1),\\
        &\tilde b_w=b_w + \boldsymbol{y}_T^H\boldsymbol{y}_T-2\Re \bigg\{\boldsymbol{y}_T^H\boldsymbol{\Phi}\boldsymbol{\mu}_h\bigg\} + \boldsymbol{\mu}_h^H\boldsymbol{H}_h\boldsymbol{\mu}_h \\
        &\quad +\operatorname{tr}\bigg\{\boldsymbol{\Sigma}_h\odot\boldsymbol{H}_h\bigg\}.
    \end{aligned}
    \label{awitr}
    \end{equation}

    The covariance matrix $\boldsymbol{\Sigma_{k_{\nu}}}$ can be reformulated by
    \begin{equation}
        \begin{aligned}
            \boldsymbol{\Sigma}_{\boldsymbol{k}_{\nu}}&=\bigg[\langle \alpha_w \rangle(\boldsymbol{\mu}_h\boldsymbol{\mu}_h^H+\boldsymbol{\Sigma}_h)\odot\boldsymbol{\Phi}_{\nu}^H(\hat{\boldsymbol{k}}_{\nu}, \hat{\boldsymbol{l}}_{\tau})\boldsymbol{\Phi}_{\nu}(\hat{\boldsymbol{k}}_{\nu}, \hat{\boldsymbol{l}}_{\tau})\\&+\rm{diag} (\langle \boldsymbol\gamma^{\nu}\rangle) \bigg]^{-1},
           \label{nusigmaitr}
        \end{aligned}
    \end{equation}
in which $\boldsymbol\gamma^{\nu}=[\gamma^{\nu}_1,\gamma^{\nu}_2,\cdots,\gamma^{\nu}_P]^T$, $\langle \gamma^{\nu}_i\rangle=\sum_{t=1}^T r_{i,t}\langle\alpha^{\nu}_t\rangle$, and $\langle\alpha^{\nu}_t\rangle={\tilde a^{\nu}_t}/{\tilde b^{\nu}_t}$. The shape parameter $\tilde{a}^{\nu}_t$ and rate parameter $\tilde{b}^{\nu}_t$ are given by
    \begin{equation}
    \begin{aligned}
        &\tilde a^{\nu}_t = a^{\nu}_t +\frac{1}{2}\sum_{i=1}^P r_{i,t},~
        \tilde b^{\nu}_t=b^{\nu}_t + \frac{1}{2}\rm{diag}(\boldsymbol{\Sigma_{k_{\nu_i}}})\boldsymbol{r}_i^T .
    \end{aligned}
    \label{alphak}
    \end{equation}
Then, mean vector  $\boldsymbol {\mu_{k_{\nu}}}$ can be given by
    \begin{equation}
        \begin{aligned}
\boldsymbol{\mu}_{\boldsymbol{k}_{\nu}}\!&=\!\boldsymbol{\Sigma}_{\boldsymbol{k}_{\nu}}\bigg[\bigg(\rm{diag}(\boldsymbol{\mu}_h^*) \boldsymbol{\Phi}_{\nu}^H(\hat{\boldsymbol{k}}_{\nu}, \hat{\boldsymbol{l}}_{\tau})\boldsymbol{y}_T-\rm{diag}(\boldsymbol{\mu}_h^*)\boldsymbol{\Phi}_{\nu}^H(\hat{\boldsymbol{k}}_{\nu}, \hat{\boldsymbol{l}}_{\tau}) \boldsymbol{\Phi}(\hat{\boldsymbol{k}}_{\nu}, \hat{\boldsymbol{l}}_{\tau})\boldsymbol{\mu}_h\\
           &+(\boldsymbol{\Sigma}_h+\boldsymbol{\mu}_h\boldsymbol{\mu}_h^H)\odot(\boldsymbol{\Phi}_{\nu}^H \boldsymbol{\Phi}_{\nu})\hat{\boldsymbol{k}_{\nu}}-{\rm diag}\big(\boldsymbol{\Phi}_{\nu}^H (\hat{\boldsymbol{k}}_{\nu}, \hat{\boldsymbol{l}}_{\tau})\boldsymbol{\Phi}(\hat{\boldsymbol{k}}_{\nu}, \hat{\boldsymbol{l}}_{\tau})\boldsymbol{\Sigma}_h\big)\bigg)\\ &\cdot\langle\alpha_w\rangle
           -\boldsymbol\gamma^{\nu}\odot\boldsymbol{\mu}^{\nu}
           \bigg],
           \label{numuitr}
        \end{aligned}
    \end{equation}
where $\boldsymbol{\mu}^{\nu}=[\mu^{\nu}_1,\mu^{\nu}_2,\cdots,\mu^{\nu}_T]^T$ can be formulated by
    \begin{equation}
    \begin{aligned}
        &\mu^{\nu}_t= {\sigma^{{\nu}}_t}^2\sum_{i=1}^P r_{i,t}l_i,~
         {\sigma^{{\nu}}_t}^2=\frac{1}{\sum_{i=1}^P r_{i,t}}.
    \end{aligned}
    \label{mul}
    \end{equation}
\begin{algorithm}[H]
    \caption{NPBL-Based Channel Estimation}
    \label{A1}
	\begin{algorithmic}
		\STATE \textbf{Input}:  $\boldsymbol{y}_T$, $P$, $\eta$, $\mathcal{T}$, $\epsilon$.  
		\STATE 1: \textbf{Initialize} $\boldsymbol{\Omega}$, $\ell =0$; Initialize $\boldsymbol{h}$, $\boldsymbol{k}_{\nu}$, $\boldsymbol{l}_{\tau}$ using the method from \cite{raviteja2019embedded}.
		\STATE 2: \textbf{repeat}
            \STATE 3: Calculate $\boldsymbol r$, $\boldsymbol\lambda_{1}$, $\boldsymbol\lambda_{2}$, $\boldsymbol{C}$ according to \eqref{ritr}, \eqref{alphahitr} and \eqref{alphaitr} .
		\STATE 4: Calculate $\boldsymbol {h}$, $\boldsymbol {k}_{\nu}$, $\boldsymbol {l}_{\tau}$ according to \eqref{hitr}, \eqref{nusigmaitr} and \eqref{numuitr}.
		\STATE 5: Update $\alpha_w$,  $\boldsymbol\alpha_h$, $\boldsymbol\alpha_l$, $\boldsymbol\alpha_k$, $\boldsymbol\mu_h$, $\boldsymbol\mu_l$ and $\boldsymbol\mu_k$ 
        according to \eqref{awitr}, \eqref{alphak} and \eqref{mul} ;
		\STATE 6: $\ell = \ell +1$;
            \STATE 7:Pruning the $i$-path if $\gamma^h_i>\eta$.
		\STATE 8: \textbf{until} $\ell\geq \mathcal{T} $  or  $\frac{||\boldsymbol{\Phi}(\boldsymbol{k}_{\nu}^{(\ell)}, \boldsymbol{l}_{\tau}^{(\ell)}) \boldsymbol{h}^{(\ell)}-\boldsymbol{\Phi}(\boldsymbol{k}_{\nu}^{\ell-1}, \boldsymbol{l}_{\tau}^{\ell-1}) \boldsymbol{h}^{\ell-1}||^2}{||\boldsymbol{\Phi}(\boldsymbol{k}_{\nu}^{\ell-1}, \boldsymbol{l}_{\tau}^{\ell-1}) \boldsymbol{h}^{\ell-1}||^2}\leq \epsilon$.
		\STATE 9:\textbf{Output}: $\boldsymbol{\hat{h}}$, $\boldsymbol{\hat{k}}_{\nu}$, $\boldsymbol{\hat{l}}_{\tau}$, $ \boldsymbol{C}$.
	\end{algorithmic}
\end{algorithm}
    
\subsection{Pruning and Convergence Condition}
The proposed algorithm learns the significance of each path through the sparsity parameter $\gamma^h_i$. A large value of $\gamma^h_i$ indicates that the corresponding path has a small gain and can be regarded as noise or a virtual path, which should be removed from the model. Therefore, a threshold $\eta$ is defined to eliminate irrelevant paths. Any path with a sparsity parameter $\gamma^h_i > \eta$ is considered negligible and is pruned to reduce model complexity and prevent overfitting.

The convergence of the algorithm is determined by the variation rate of the reconstructed signal between two consecutive iterations. The convergence criterion at the $i$-th iteration is defined as:
\begin{equation}
   \Delta_i= \frac{||\boldsymbol{\Phi}(\boldsymbol{k}_{\nu}^{(t)}, \boldsymbol{l}_{\tau}^{(t)}) \boldsymbol{h}^{(t)}-\boldsymbol{\Phi}(\boldsymbol{k}_{\nu}^{i-1}, \boldsymbol{l}_{\tau}^{i-1}) \boldsymbol{h}^{i-1}||^2}{||\boldsymbol{\Phi}(\boldsymbol{k}_{\nu}^{i-1}, \boldsymbol{l}_{\tau}^{i-1}) \boldsymbol{h}^{i-1}||^2}.
\end{equation}

The iterative process terminates when $\Delta^{(i)} \leq \epsilon$  or when the number of iterations $i$ reaches $\mathcal{T}$.  
All the above procedures are summarized in Algorithm~\ref{A1}. {The method updates posterior distributions iteratively, involving large-scale matrix inversion and multiplication operations. The overall computational complexity can be summarized as $\mathcal O\big(P^3 +(l_{\max}+1)(2k_{\max}+1)P^2\big)$. The algorithm incorporates the pruning mechanism thus lowering computational overhead while preserving estimation accuracy.}  

\begin{table}[H]
		\centering
		\caption\centering{Simulation Parameters}
		\setlength{\tabcolsep}{20pt} 
		\label{SIMULATIONPARAMETERS}
            \normalsize
		\begin{tabular*}{\linewidth}{ll}
		\midrule
                 Parameter & Value \\
            \midrule
                 The size of DD grid  & $M$=32, $N$=32 \\
                 Subcarrier spacing  & 15 kHz \\
                Carrier frequency & 5.9 GHz \\
                Maximum delay & $8.3\times 10^{-6} s$ \\
                Maximum velocity & 500 km/h\\
                Pilot position  & $l_p=16, k_p=16$ \\
       	\midrule
    \end{tabular*}
\end{table}

\section{Simulation Results}

In this section, we compare the performance of the proposed NPBL scheme with three existing benchmark schemes, i.e., the SBL~\cite{zhao2020sparse}, 1D SBL-offgrid~\cite{wei2022off}, and VB-GMM~\cite{kong2021variational}. More specifically, we consider the resulting normalized mean squared error (NMSE) of the estimated channel coefficients, defined by 
\begin{equation}
    {\text{NMSE}} =  {\frac{{\left. {\mathbb{E} \bigg(\left\| {{\boldsymbol{H}} - \widehat {\boldsymbol{H}})} \right\|_2^2} \right)}}{{\mathbb{E}\left( {\left\| {\boldsymbol{H}} \right\|_2^2} \right)}}}.
\end{equation}
We consider a high-speed train application scenario where the OTFS system parameters are summarized in Table~\ref{SIMULATIONPARAMETERS}. The algorithmic parameters are designed as $\lambda_1=1$, $\lambda_2=0.05$, $\boldsymbol a_h =\boldsymbol a_l =\boldsymbol a_k = 10^{-3} \cdot \mathbf{1}^{1 \times T}$, $\boldsymbol{b}_h = \boldsymbol b_l =\boldsymbol b_k = \mathbf{1}^{1 \times T}$, with $c = d = 10^{-2}$, $\mathcal{T} = 500$, and $\eta = 10^{3}$. 
A simplified CDL model is adopted, where each cluster contains $3-5$ randomly generated propagation paths. While reduced in complexity compared to the standard 20-ray model, it retains basic multipath structure and is suitable for comparative studies of estimation algorithms. For simplicity, the full frame is utilized for channel estimation in our simulations. 

\begin{figure}[ht]
        \centering
        \includegraphics[width=0.7\linewidth]{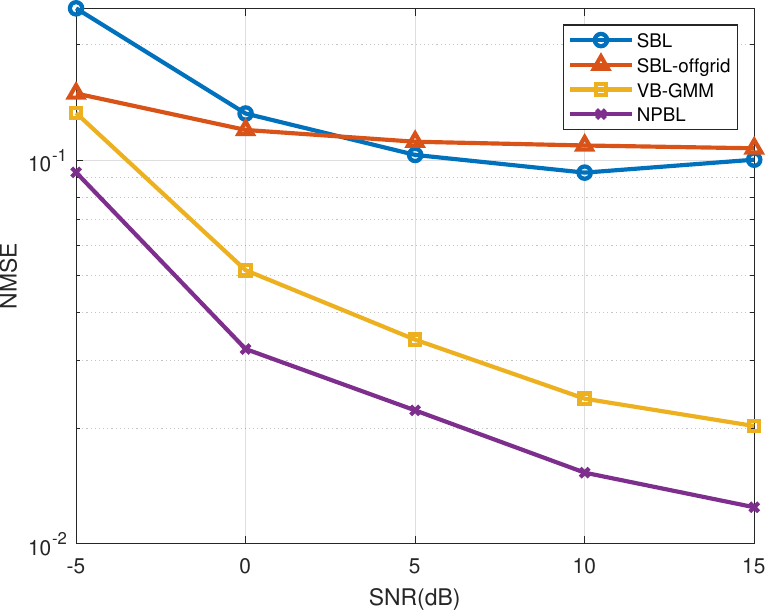}
        \caption{NMSE of SNR for different algorithms.}
        \label{NMSESNR}
    \end{figure}

In Fig.~\ref{NMSESNR}, we compare the NMSE performance of the proposed NPBL method against three existing schemes across a range of SNRs from -5 dB to 15 dB under the condition of 3 clusters in the channel. The results demonstrate that NPBL consistently achieves the lowest NMSE, particularly in medium to high SNR regimes. This superior performance can be attributed to NPBL’s ability to its inherent capability for automatic cluster classification and parameter estimation, thereby overcoming a key limitation of standard SBL. Furthermore, NPBL excels at modeling complex channel distributions in a non-parametric manner, outperforming even advanced methods like VB-GMM, which relies on a fixed Gaussian mixture structure. By flexibly learning arbitrary distributions without being constrained by pre-defined parametric assumptions, NPBL adapts more accurately to realistic channel conditions.
    
\begin{figure}[ht]
        \centering
        \includegraphics[width=0.7\linewidth]{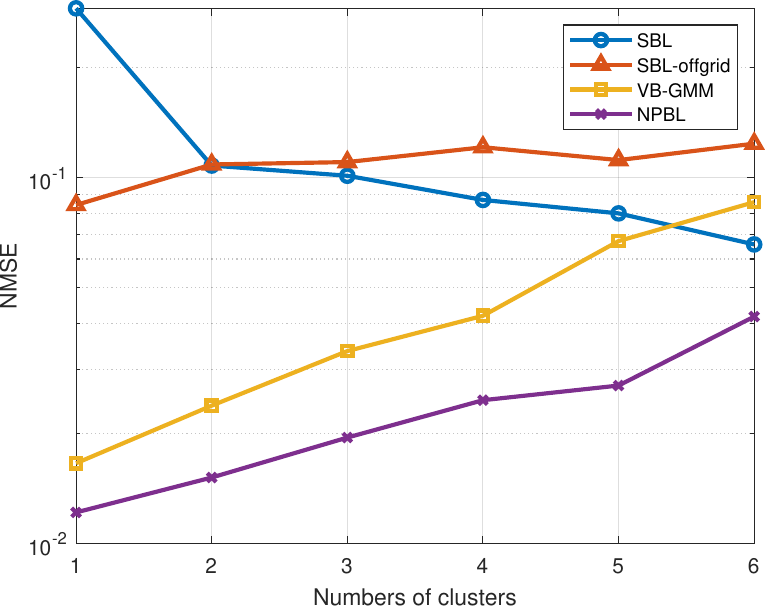}
        \caption{NMSE of SNR for different cluster numbers.}
        \label{Numbers of clusters}
    \end{figure}

We illustrate the impact of cluster number on NMSE performance at SNR = 10 dB, highlighting NPBL’s robustness and consistency compared to SBL, SBL-offgrid, and VB-GMM in Fig.~\ref{Numbers of clusters}.  
The results show the superior model order selection capability of NPBL and its robustness to variations in channel sparsity and cluster complexity. Compared to VB-GMM, which is limited by its parametric Gaussian mixture form, NPBL’s non-parametric nature enables it to infer cluster structures and path dependencies directly from the data, ensuring accurate estimation even when the number of clusters changes.

\section{Conclusion}
We propose an NPBL framework for channel estimation with OTFS in both SNR-varying and cluster-varying scenarios. Our design demonstrated significant advantages over existing techniques since the non-parametric Bayesian structure provides greater flexibility and accuracy in modeling realistic wireless channels.
These results showed that NPBL is not only a more accurate estimator but also a more robust and versatile solution, making it highly suitable for practical OTFS and multi-antenna systems where channel conditions are dynamic and complex. This capability is particularly valuable for SAGIN in future 6G communications, where high mobility, diverse propagation conditions, and dynamic cluster behaviors are prevalent. Future work may focus on further reducing computational complexity while maintaining these performance benefits, thereby supporting the scalable deployment of NPBL in next-generation communication systems.
\balance
\bibliographystyle{ieeetr}
\bibliography{ref}
\end{document}